\documentclass[aps,prl,reprint,amsmath,amssymb,superscriptaddress,showpacs,bibnotes]{revtex4-1}

\usepackage{graphicx}
\usepackage{color}
\newcommand\fwo{\textsc{fwo}}
\newcommand\bwo{\textsc{bwo}}
\newcommand\fwc{\textsc{fwc}}
\newcommand\bwc{\textsc{bwc}}
\newcommand\spin{\textsc{spin}}
\newcommand\s{\textsc{s}}
\newcommand\oo{\textsc{o}}
\newcommand\rc{\textsc{c}}
\newcommand\fw{\textsc{fw}}
\newcommand\bw{\textsc{bw}}

\begin{document}

\title{Coherent and Dynamic Beam Splitting based on Light Storage in Cold Atoms}

\author{Kwang-Kyoon Park}
\affiliation{Department of Physics, Pohang University of Science and Technology (POSTECH), Pohang 37673, Korea}
%\email{Kwangkyoon.Park@gmail.com}

\author{Tian-Ming Zhao}
\affiliation{Department of Physics, Pohang University of Science and Technology (POSTECH), Pohang 37673, Korea}

\author{Jong-Chan Lee}
\affiliation{Department of Physics, Pohang University of Science and Technology (POSTECH), Pohang 37673, Korea}

\author{Young-Tak Chough}
\affiliation{Department of Medical Technology, Gwangju University, Gwangju 61743, Korea}

\author{Yoon-Ho Kim}
\email{yoonho72@gmail.com}
\affiliation{Department of Physics, Pohang University of Science and Technology (POSTECH), Pohang 37673, Korea}

\date{November 16, 2015}

%%%%%%%%%%%%%%%%%%%%%%%%%%
\begin{abstract}
We demonstrate a coherent and dynamic beam splitter based on light storage in cold atoms. An input weak laser pulse is first stored in a cold atom ensemble via electromagnetically-induced transparency (EIT). A set of counter-propagating control fields, applied at a later time,  retrieves the stored pulse into two output spatial modes. The high visibility interference between the two output pulses clearly demonstrates that the beam splitting process is coherent. Furthermore, by manipulating the control lasers, it is possible to dynamically control the storage time, the power splitting ratio, the relative phase, and the optical frequencies of the output pulses. The active beam splitter demonstrated in this work is expected to  significantly reduce the resource requirement in photonic quantum information and in all-optical information processing as a single cold atom ensemble can functionally replace a variety of optical elements, including beam splitters, mirrors, phase shifters, and optical quantum memories.
\end{abstract}
%%%%%%%%%%%%%%%%%%%%%%%%%%

% Long abstract %
%We demonstrate a coherent and dynamic beam splitter by means of light storage in cold atoms. An input weak laser pulse is first stored in a cold atom ensemble via electromagnetically-induced transparency (EIT). A set of counter-propagating control fields, applied at a later time,  retrieves the stored pulse into two output spatial modes, completing the beam splitting operation. The high visibility interference of $99.1\%$ between the two output pulses clearly demonstrates that the beam splitting process is coherent. Furthermore, by manipulating the control lasers, it is possible to dynamically control the power splitting ratio, the relative phase between the two output pulses, and the optical frequencies of the output pulses. Since our beam splitter is based on EIT light storage in cold atoms, it is also possible to dynamically manipulate the storage time, i.e., the time delay between the input and the outputs. The coherent and dynamic beam splitter demonstrated in this work is expected to  significantly reduce the resource requirement in photonic quantum information processing as well as in all-optical information processing as a single cold atom ensemble can functionally replace a variety of optical elements, including beam splitters, mirrors, phase shifters, and optical quantum memories.

%(1) Choose no more than four index number codes. (2) Place your principal index code first.
%(3) Always choose the lowest-level code available. (4) Always include the check characters.

%32.80.Qk	Coherent control of atomic interactions with photons
%42.50.Ex	Optical implementations of quantum information processing and transfer
%42.50.Gy	Effects of atomic coherence on propagation, absorption, and amplification of light; electromagnetically induced transparency and absorption

\maketitle

Photonic quantum computation has received much attention over the last decade due to the demonstration of scalability using single-photon sources, projective measurement, and linear optical elements such as beam splitters, phase shifters, and mirrors \cite{Knill:2001is}. In addition to these elements,  quantum memories are essential in synchronizing  multi-photon events to increase the success probability of linear optical quantum gates \cite{Knill:2001is,Kok:2007ep,Bussieres:2013br}. Recent advances in miniaturized photonic circuits on a silicon chip may reduce the experimental overhead in managing many such linear optical elements \cite{Carolan2015}. Here, we demonstrate a cold-atom based approach to consolidate different linear optical operations and a quantum memory in a single device. By employing EIT-based light storage, we demonstrate that a weak laser pulse can be coherently and dynamically manipulated in a cold atom ensemble to perform beam splitting, reflection, and phase shifting, in addition to quantum memory operation.
%Our work has important implications in resource-efficient photonic quantum information processing and all-optical information processing. 
%}

% quantum memory
% linear optical device = beam splitter, phase shifter, mirror 

Beam splitters play a crucial role in quantum optics and photonic quantum information, as demonstrated in the Hanbury-Brown--Twiss experiment \cite{HanburyBrown1956}, the Hong-Ou-Mandel effect \cite{Hong:1987}, implementation of photonic quantum gates \cite{Knill:2001is,Carolan2015}, etc. While  beam splitters based on passive linear optics are widely available at low cost, there are growing needs  for dynamic beam splitting devices in which the splitting ratio can be easily varied. Moreover, such devices integrated with coherent optical memory functionalities can significantly reduce resource requirement in photonic quantum information and all-optical information processing. Coherent optical memories or quantum memories for photons have recently been demonstrated by utilizing coherent atom-photon interaction schemes, e.g., EIT \cite{Fleischhauer:2000vm,Fleischhauer:2002ee,Phillips:2001ec,Liu:2001hu}, the DLCZ protocol \cite{Zhao:2008kz,Zhao:2008hz}, and Raman memory \cite{Sprague:2014,Ding:2015}. Dynamic beam splitting devices based on such atom-photon interaction schemes, therefore, would allow an integrated photonic device in which quantum memory and linear optical functionalities are combined.      

Recently, controllable beam splitting in frequency or time modes has been demonstrated by using the gradient echo memory scheme \cite{Campbell:2012ew}, with a double-tripod EIT scheme \cite{Lee:2014}, and with a tripod EIT scheme \cite{Yang:2015ce}. Dynamic beam splitting in the spatial modes which functions as a variable beam splitter \cite{Baek:2008}, however, has not been implemented in a long-coherence atomic medium. Spatial beam splitting has, nevertheless, been demonstrated in warm atomic vapor relying on the EIT effect and natural atomic thermal motions in which atoms with light-induced ground state coherence can move freely to another spatial channel at which the ground state coherence is retrieved back to light  \cite{Xiao:2008ba}. Since this scheme relies on atomic thermal random motion rather than a phase-matched process, only a small fraction of the input light can be retrieved at the second spatial channel. There have been several theoretical proposals for efficient spatial splitting of a signal beam by using a cold atom medium with two control beams  \cite{Hansen:2007hr,Hansen:2007hq,Nikoghosyan:2009dx}, but due to different physics, these schemes are inapplicable for warm atomic vapor. For instance, in warm vapor, light retrieval can occur only in the spatial mode with the stronger control field \cite{Zimmer:2006fw}, whereas, in the cold atoms, the input signal beam is split into two spatial modes  according to the power ratio of the two control fields \cite{Hansen:2007hr}.

%%%%%%%%%%%%%%%%%%%%%%%%%%%%%%%%%%%%%%%%%%
\begin{figure*}[tp]
\centering
\includegraphics[width=6.50 in]{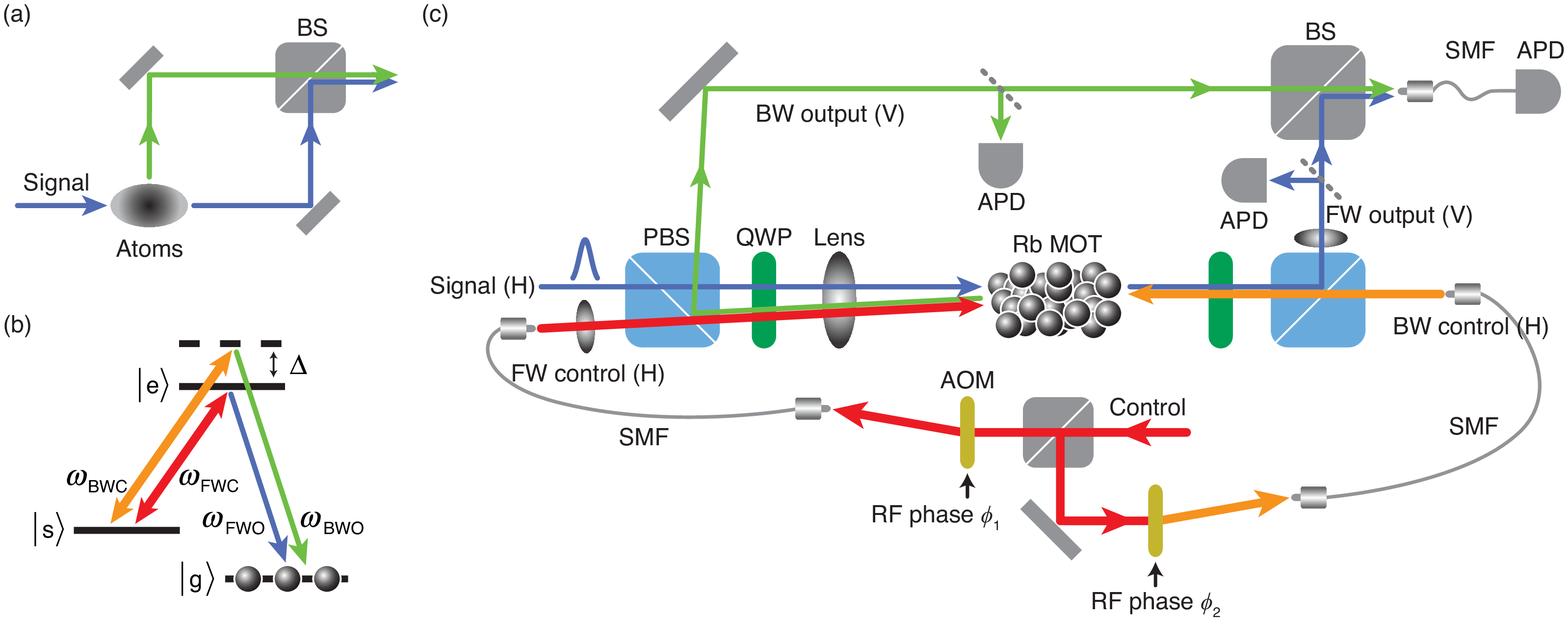}
\caption{Experimental setup. (a) Conceptual schematic of the Mach-Zehnder interferometer utilizing  the  beam splitter based on EIT light storage in a cold atom ensemble.
(b) The input signal  is resonant with $| g \rangle$-$| e \rangle$. Two control lasers are applied between $|s\rangle$-$|e\rangle$: on resonant for the forward control ($\omega_\fwc$) and with a detuning $\Delta$ for the backward control ($\omega_\bwc$). 
The input signal is then split into the two output beams: forward output ($\omega_\fwo$, resonant to $| g \rangle$-$| e \rangle$)  and backward output ($\omega_\bwo$, with a detuning $\Delta$ to $| g \rangle$-$| e \rangle$). All fields are $\sigma^+$ polarized. 
(c) The weak signal pulse is launched into an $^{87}$Rb MOT when the strong forward (FW) control beam is turned on. The angle between the FW control and the signal is $0.3^\circ$. Storage of the signal is achieved simply by turning off the FW control. After a certain storage time, the input signal can be retrieved into the forward (FW) or backward (BW) outputs by turning on the FW or BW controls, respectively. AOM: acousto-optic modulator. BS: non-polarizing beamsplitter. PBS: polarizing beamsplitter. QWP: quarter wave plate. SMF: single-mode fiber. APD: avalanche photodiode.
}
\label{scheme}
\end{figure*}
%%%%%%%%%%%%%%%%%%%%%%%%%%%%%%%%%%%%%%%%%%

In this paper, we propose and demonstrate coherent and dynamic beam splitting of a weak laser pulse into two spatial modes, precisely resembling a variable beam splitting operation of a linear optical device \cite{Baek:2008}.  Our scheme is based on the EIT-based light storage scheme in a cold atom ensemble: The photonic excitation of a weak laser pulse is mapped into an atomic coherence, called the spin wave \cite{Fleischhauer:2000vm,Fleischhauer:2002ee}. After a controlled delay, two nearly counter-propagating control beams are applied to retrieve the signal beam into two spatial modes whose directions are determined by the phase matching condition. As the scheme is based on a phase-matched process, light retrieval can be highly efficient. Experimentally, we demonstrate dynamic beam splitting with the quantum memory operation. We show that, between the two output spatial modes, it is possible to control the power splitting ratio, the relative phase, and the relative frequency shift. High visibility interference between the two output modes confirms that the beam splitting process is coherent.

%%%%%% Main Body %%%%%%%%%%
The essential features of the experiment are depicted in Fig.~\ref{scheme}(a). A weak signal pulse of $\sim$50 nW is split into two directions at the atomic ensemble after a certain period of coherent light storage. The two output beams are then superposed at a linear optical beam splitter to form a Mach-Zehnder interferometer. The energy level diagram for the cold atoms and the detailed experimental schematic are shown in Fig.~\ref{scheme}(b) and Fig.~\ref{scheme}(c), respectively. 87 rubidium cold atoms ($< 1$ mK) are prepared by a magneto-optical trap (MOT). After loading the neutral atomic ensemble, the trapping magnetic field is turned off and all the atoms are optically pumped into the ground state, $\left| g \right\rangle  = \left| {5{S_{1/2}},F = 1,m_F = 1} \right\rangle$. The other relevant atomic levels are $\left| e \right\rangle  = \left| {5{P_{1/2}},F' = 2,m_{F'} = 2} \right\rangle$  and $\left| s \right\rangle  = \left| {5{S_{1/2}},F = 2,m_F = 1} \right\rangle $. The quantization axis is set to be the same as the propagation axis of the signal field. The properties of the medium is characterized by measuring the two-level transmission and the three-level EIT transmission spectrum. Optical depth (OD) is measured to be 21 and the ground state dephasing rate is measured to be $\gamma_{gs} = 2\pi \times 3.8{\text{ kHz}}$. The decay rate of the $\left| g \right\rangle$-$\left| e \right\rangle$ transition is $\Gamma/2 = 2\pi \times 2.9{\text{ MHz}}$. The forward (FW) control laser is first turned on, illuminating the whole atomic ensemble, after which the signal pulse is sent for the beam splitting experiment. The whole procedure is repeated every 50 ms.

The horizontally (H) polarized input signal pulse is transformed to right circular polarization and focused into the atomic ensemble of cold atoms. The FW control is turned on to open the transparency window for the signal pulse and then adiabatically turned off to map the signal pulse into the atomic spin wave. The angle between the FW control and the signal is $0.3^\circ$. After a programmable storage, the controls are turned back on to convert the spin wave back into photonic excitation, i.e., the output beams. The  direction ${\vec k_\oo}$ and the frequency ${\omega _\oo}$ of the output are determined by the following phase matching condition \cite{Zibrov:2002if},
\begin{align}
{\vec k_\s} - {\vec k_\fwc} = {\vec k_\spin} =  {\vec k_\oo} - {\vec k_\rc}, \label{phasek}\\
{\omega _\s} - {\omega _\fwc} ={\omega _\spin} =  {\omega _\oo} - {\omega _\rc}, \label{phasew}
\end{align}
where ${\vec k_\spin}$ (${\vec k_\s}$, ${\vec k_\fwc}$) and ${\omega _\spin}$ (${\omega _\s}$, ${\omega _\fwc}$) are the wave vector and frequency of the spin wave (signal, FW control), respectively. The wave vectors and the frequency of the control beams (i.e., retrieval control) used for generating the output beams are ${\vec k_\rc}$ and ${\omega _\rc}$, respectively.

%%%%%%%%%%%%%%%%%%%%%%%%%
\begin{figure}[tp]
\centering
\includegraphics[width=3.4 in]{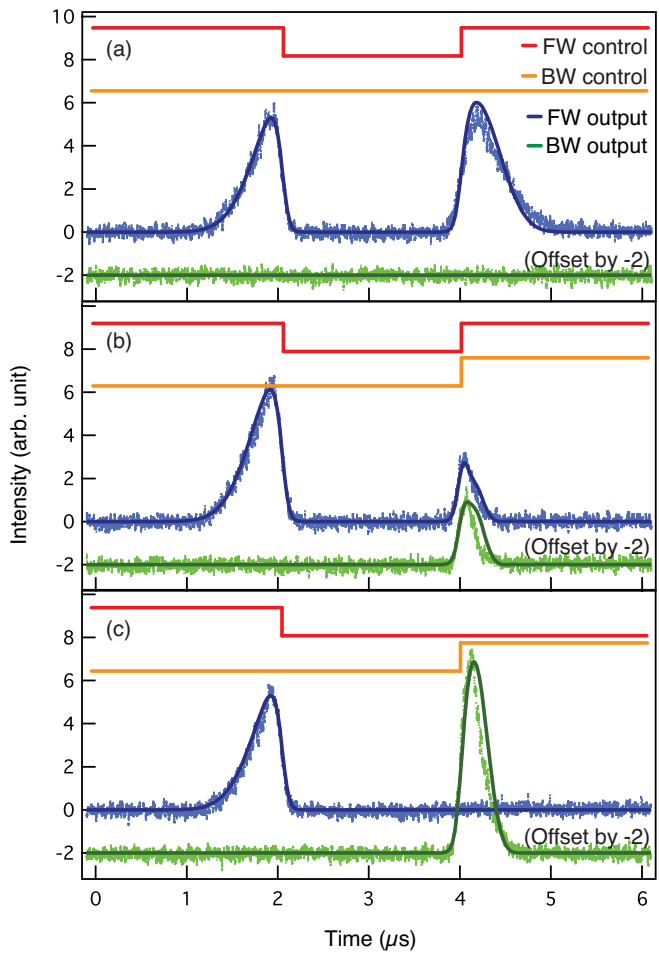}
\caption{Variable beam splitting. The thick red (orange) line represents the on/off sequence for the FW (BW) control laser. The input signal is stored into the atomic ensemble by turning off the FW control laser. The blue (green) trace shows the detected  FW (BW) output.  
(a) 100:0 beam splitting. The stored signal is retrieved completely into the FW output by turning on the FW control. 
(b) 50:50 beam splitting. The stored signal is retrieved into both the FW and BW outputs by simultaneously applying the FW and BW controls at the same Rabi frequency. 
(c) 0:100 beam splitting. The stored signal is retrieved completely into the BW output by turning on the BW control. In all cases, $\Delta = 0$. The superimposed  solid curves are due to numerical simulation. See text for details.}
\label{splitting}
\end{figure}
%%%%%%%%%%%%%%%%%%%%%%%%%
%  at 2.05 $\mu$s
%  After a controllable delay about 1.93 $\mu$s

According to Eq.~(\ref{phasek}), if we assume $\vec k_\rc = \vec k_\fwc$, then $\vec k_\oo = \vec k_\s$,  i.e., the direction of the output signal is the same as that of the input if the retrieval control is the same as the FW control. On the other hand, if we consider the condition $\vec k_\rc \simeq -\vec k_\s$, then $\vec k_\oo \simeq - \vec k_\fwc$. That is, if  the retrieval control is in the backward (BW) direction, the output is nearly opposite to the FW control as sketched in Fig.~\ref{scheme}(c). The input signal beam can be split into two output directions if both FW and BW controls are applied to the cold atom medium at the same time, thus functioning as a dynamic beam splitter. Note that this process is not possible in a warm vapor. In a medium with a large Doppler shift, instead of beam splitting, light retrieval can occur only in the spatial mode with the stronger control field. This is due to the fact that higher spatial-frequency components of Raman coherence decay quickly in warm vapor \cite{Hansen:2007hr,Zimmer:2006fw}.

% The right circular polarization of the outputs are transformed to vertical (V) polarization and collimated with a lens. 

%%%%%%%%%%%%%%%%%%%%%%%%%
\begin{figure}[tp]
\centering
\includegraphics[width=3.4 in]{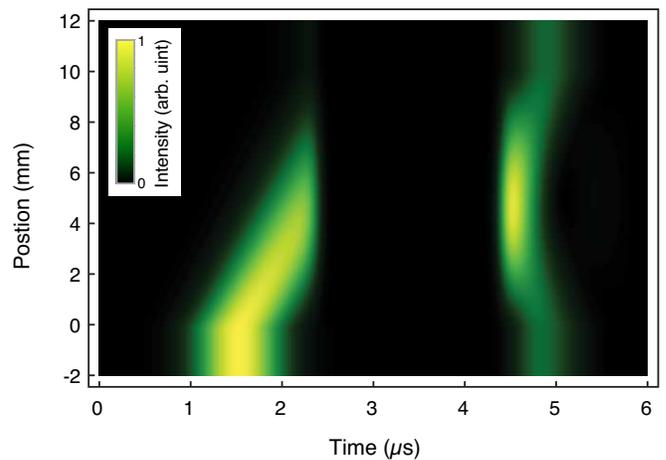}
\caption{Numerical simulation of the signal propagation. The group velocity of the signal is reduced when it encounters the EIT medium located between 0 to 10 mm. The photonic component of the signal disappears at 2.4 $\mu$s when the FW control is turned off as it is mapped to the atomic spin wave excitation. By turning on the FW and BW controls at 4.4 $\mu$s, the input signal stored in the EIT medium is retrieved into the FW and BW outputs.  OD is set to be 100 to consider a situation with negligible signal leakage.    
}
\label{simul}
\end{figure}
%%%%%%%%%%%%%%%%%%%%%%%%%

The experimental results for dynamic beam splitting are shown in Fig.~\ref{splitting}. Here, the detuning $\Delta$ of the BW control from the $\left| s \right\rangle$-$\left| e \right\rangle$ transition  is set to zero. Part of the input pulse is stored in the cold atom medium by turning off the FW control. The Rabi frequency of the FW control is ${\Omega _\fwc} = 2\pi  \times (5.8 \pm 0.2) {\text{ MHz}}$. The early-arriving part of the pulse leaks out from the medium without being stored before the turn-off of the FW control at 2.05 $\mu$s. After a controllable delay which is set to 1.93 $\mu$s, the controls are turned on for mapping the spin wave back into photonic excitation. In Fig.~\ref{splitting}(a), only the FW control is turned on at 3.98 $\mu$s with ${\Omega _\fwc} = 2\pi  \times 5.7{\text{ MHz}}$. Just as in normal EIT storage, all the stored light is retrieved into the FW output, making the cold atom ensemble to function as a coherent optical buffer or a quantum memory. For 50:50 beam splitting, the BW control with ${\Omega _\bwc} = 2\pi  \times 6.6{\text{ MHz}}$ is simultaneously turned on with the FW control as shown in Fig.~\ref{splitting}(b). In this case, the spin wave is split into almost opposite directions in the atomic ensemble and the two counter-propagating output beams are retrieved. The medium thus functions as a 50:50 beam splitter if the effective strengths of the FW and BW controls are the same. If effective Rabi frequencies of the controls are different, the medium functions as an asymmetric beam splitter in which more power is emitted in the direction of the stronger control. As shown in Fig.~\ref{splitting}(c), it is possible that all the stored light is retrieved in the backward direction by turning on the BW control only, thus making the medium to function as a mirror integrated with a quantum memory. The pulse width of the BW output in Fig.~\ref{splitting}(c) is narrower than that of Fig.~\ref{splitting}(a) due to the fact that the BW control is larger, ${\Omega _\bwc} = 2\pi  \times 7.7{\text{ MHz}}$ for Fig.~\ref{splitting}(c).

%%%%%%%%%%%%%%%%%%%%%%%%%%%%
\begin{figure}[tp]
\centering
\includegraphics[width=3.4 in]{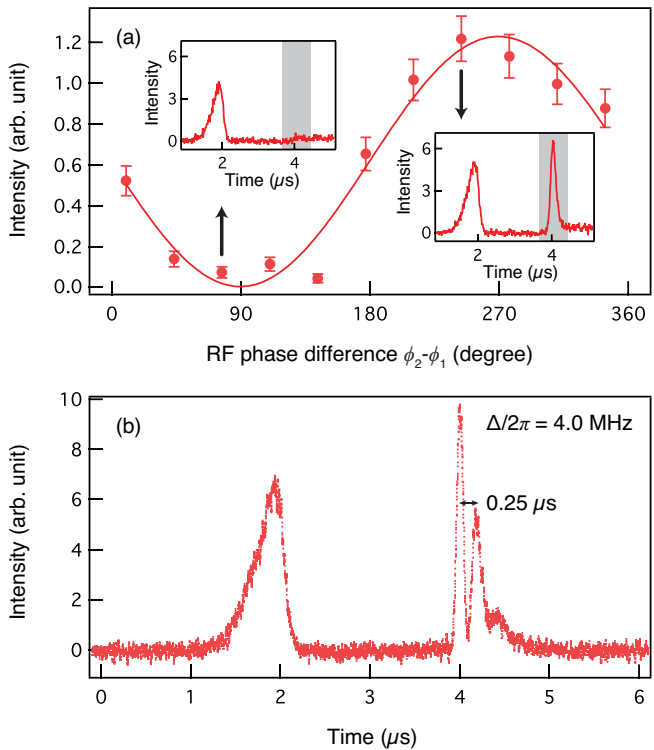}
\caption{Demonstration of coherence between the two ouputs. 
(a) High visibility interference of $99.1\%$ is observed between the two output beams in the experimental setup depicted Fig.~\ref{scheme}. The phase difference between the two output beams can be controlled by changing the RF phase difference $\phi_2-\phi_1$. Each data point is acquired by integrating the grey region of the output trace shown in the inset.  The experiment is done at $\Delta  = 0$. 
(b) If the detuning $\Delta$ is non-zero for the BW control, the frequencies of the BW and FW outputs will differ, thereby causing beating between the two outputs. 
}
\label{interference}
\end{figure}
%%%%%%%%%%%%%%%%%%%%%%%%%%%%
% Here, by setting the detuning $\Delta$ at $\Delta/2\pi = 4.0{\text{ MHz}}$, beating between the two outputs is observed at the oscillation period of 0.25 $\mu$s. 

Our experimental results are well explained by the numerical simulation shown in the solid lines of Fig.~\ref{splitting}. The numerical simulation is based on the Maxwell-Bloch equations for light propagation \cite{Lin:2009}. By comparing the experimental data in Fig.~\ref{splitting} and the results of numerical simulation using the measured values of OD and $\gamma_{gs}$, we extract the effective values of the Rabi frequencies. The two-dimensional numerical simulation in Fig.~\ref{simul} showing the singal propagation in the medium was done by further assuming ${\Omega _\fwc} = {\Omega _\bwc} = 2\pi  \times 7.6{\text{ MHz}}$ and OD is set to be 100 to consider a situation with negligible signal leakage.  

% The signal pulse propagating at the vacuum speed of light slows down as it enters the EIT medium located from 0 to 10 mm. At 2.4 $\mu$s, the light component is fully mapped onto the spin wave in the cold atom ensemble. After applying the two two counter-propagating retrieval controls at 4.4 $\mu$s, the stored light is retrieved into two directions and escape the medium into opposite directions.

%\subsection{Interference.}
We now examine coherence between the two output beams by superposing them at a linear optical beam splitter, forming a Mach-Zehnder interferometer  in which the first beam splitter is made of a cold atom ensemble, see Fig.~\ref{scheme}(a). If the FW and BW output beams in Fig. \ref{splitting}(b) are mutually coherent, high visibility interference is expected. For this experiment, the detuning $\Delta$ is set to zero so that the frequencies of the FW and BW outputs are identical. The experimental result is shown in Fig.~\ref{interference}a and it clearly shows high visibility interference between the FW and BW output beams, confirming that the beam splitting process is coherent.
 It is interesting to note that the phase difference between the FW and BW output beams can be varied by changing the relative phase between the two control fields. In experiment, the relative phase is controlled by the phase difference between the RF signals, $\phi_2 - \phi_1$, applied to the acousto-optic modulators (AOM) for generating the FW and BW controls. This results  suggests that, with the input signal in the single-photon state, our scheme can be used to prepare a single-photon dual-rail entangled state of the form,
%\begin{equation*}
$
\left| \psi  \right\rangle  = \sin \theta {\left| 1 \right\rangle _\fw}{\left| 0 \right\rangle _\bw} + \cos \theta {e^{i\delta \phi }}{\left| 0 \right\rangle _\fw}{\left| 1 \right\rangle _\bw},
$
%\end{equation*}
where $\tan \theta  = {\Omega _\fwc}/{\Omega _\bwc}$ and $\delta \phi  = {\phi _2} - {\phi _1}$.

% $\delta \phi  = {\phi _\bwc} - {\phi _\fwc}$.

As seen from Eq.~(\ref{phasew}), it is possible to perform a dynamic beam splitting operation in which one of the two outputs has slightly different frequency than that of the input if $ \omega_\rc \neq \omega_\fwc$. In experiment, this can be achieved by simply setting $\Delta  \ne 0$ in Fig.~\ref{scheme}(b). The experimental result for such a case is shown in Fig.~\ref{interference}(b) in which beating is observed with the period of $2\pi /\Delta$. This result reconfirms that the dynamic and coherent nature of the beam splitting operation based on the light storage in cold atoms.

In conclusion, we have demonstrated a coherent and dynamic beam splitter by means of an EIT-based light storage in cold atoms. Since coherent conversion between photonic and  atomic spin wave excitations is at the heart of the atom-based optical beam splitter, our beam splitter scheme naturally incorporates quantum memory functions. If, instead of the  conventional EIT medium, the Rydberg-EIT medium is used \cite{Dudin:2012hm,Peyronel:2012je}, our beam splitter scheme can further incorporate the process of single-photon filtering from a weak laser, potentially enabling direct generation of single-photon dual-rail entanglement from a weak laser. Although the beam splitter scheme was demonstrated with a weak laser pulse, it is nevertheless applicable to the single-photon input signals \cite{Choi:2008el,Zhang:2011bk}. Our scheme may also be used for other types of EIT media with negligible Doppler broadening \cite{Hansen:2007hr}. Note that, in this work, we have demonstrated a $1\times2$ beam splitting device based on a cold atom medium. It is however possible to extend the scheme to $N \times M$ beam splitting, supporting $N$ input and $M$ outputs, using the fact that the EIT medium can be excited with multiple spin wave components and the phase matching condition allows various possible directions for light storage and retrieval. Such an $N\times M$ beam splitter based on the EIT medium may offer new possibilities for multiplexed all optical information and photonic quantum information.

This work was supported by Samsung Science \& Technology Foundation under Project Number SSTF-BA1402-07.

%%%%%%%%%%%%%%%%%%%%%%%%%%%%%%%%

%\vspace*{3mm}
%\textbf{Methods}

%%%%%%%%%%%%%%%%%%%%%%%%%%%%%%%%%%%%%%

%%%%%%%%%%%%%%%
%%%%%%%%%%%%%%%

%\newpage
%
%%\vspace*{3mm}
%%\textbf{Acknowledgements}
%\section{Acknowledgements}
%This work was supported by Samsung Science \& Technology Foundation under Project Number SSTF-BA1402-07.
%
%
%%\vspace*{3mm}
%%\textbf{Author contributions}
%\section{Author contributions}
%K.-K.P. and Y.-H.K. conceived the idea and designed the experiment. K.-K.P.,T.-M.Z, and J.-C.L. performed the experiment. Y.-T.C. carried out numerical simulations. All authors discussed and analyzed the results, and contributed to writing the manuscript. All stages of the work were supervised by Y.-H.K.
%
%%\vspace*{3mm}
%%\textbf{Additional information} 
%
%\section{Additional information}
%The authors declare no competing financial interests. Correspondence and requests for materials should be addressed to Y.-H.K. (yoonho72@gmail.com).

\end{document}